# Influence of length and measurement geometry on magnetoimpedance in $La_{0.7}Sr_{0.3}MnO_3$

A. Rebello and R. Mahendiran[*]

Department of Physics and NUS Nanoscience & Nanotechnology Initiative (NUSNNI), Faculty of Science, National University of Singapore, 2 Science Drive 3, 117542, Singapore

**Abstract**

We show that ac magnetoresistance at room temperature in $La_{0.7}Sr_{0.3}MnO_3$ is extremely high ($\approx$ 47% in $\mu_0 H$ = 100 mT, $f$ = 3-5 MHz), and magnetic field dependence of reactance exhibits a double peak behavior. However, magnitudes of the ac magnetoresistance and magnetoreactance for a fixed length of the sample ($l_i$) decrease with decreasing separation ($l_v$) between voltage probes unlike the dc magnetoresistance. On the contrary, change in $l_i$ has a negligible influence on magnetoimpedance when $l_v$ is fixed. Our results indicate that high frequency electrical transport is sensitive to local variations in the magnetic permeability.

---

[*] Corresponding author – phyrm@nus.edu.sg

Perovskite manganites, of the general formula $R_{1-x}B_x MnO_3$ ($R$ = trivalent rare earth ion, $B$ = divalent alkaline earth ion), have spurred a lot of research excitement in the past one decade owing to their spectacular magnetoresistive responses, namely colossal magnetoresistance (CMR), and related technological impact as sensors[1]. While a great deal of study on the CMR in these manganese oxides have been carried out, the technological exploitation of the CMR effect at room temperature is hampered by the need of a very high magnetic field, $\mu_0 H > 1$ T to induce more than 10 % magnetoresistance (MR) measured with direct current. The low field magnetoresistance, which is attributed due to the spin polarized tunneling of electrons between ferromagnetic grains, is maximum at the lowest temperature but becomes negligible at the ferromagnetic to paramagnetic transition temperature ($T_C$). Interestingly, radio frequency (rf) magnetotransport ($f$ = 1-30 MHz) provides an alternative strategy to enhance the low field magnetoresistance at room temperature.[2,3,4,5]

The magnetic skin depth ($\delta_m = \sqrt{2/(\omega\sigma\mu_0\mu_\phi)}$) of a ferromagnetic conductor of conductivity σ, carrying an alternating current (ac) of frequency $\omega$, dramatically increases under an axial dc magnetic field due to the suppression of the ac circumferential permeability $\mu_\phi$. As a result, the impedance decreases dramatically at low magnetic fields. This phenomenon, aptly known as the giant magnetoimpedance (MI), was initially discovered in amorphous soft ferromagnetic wires and ribbons.[6] The study of MI offers additional advantage, over the static measurements, in elucidating the dynamical magnetization processes in the samples through the frequency and field dependence of $\mu_\phi$. For instance, a few studies on electromagnetic absorption in the microwave and radio frequency range, using techniques such as cavity resonance,[7,8] tunnel diode,[9] and rf oscillator[10] techniques demonstrated additional features that were not clearly visible in the dc resistivity and dc magnetization measurements.

In the present work, we show that the ac magnetoresistance in $La_{0.7}Sr_{0.3}MnO_3$ (LSMO) at room temperature is extremely high ($\approx$ 47 % in $\mu_0H$ = 100 mT and for $f$ = 3-5 MHz), and the field dependence of the reactance exhibits a double peak behavior. In addition, we show how the magnitude of the magnetoresistance and the features in the magnetoreactance are affected by the measurement geometry in LSMO. We have performed two types of distance variation: (i) the distance between the voltage probes ($l_v$) is varied while keeping the distance between the current electrodes ($l_i$) constant and (ii) The distance between voltage probes is fixed, but the length of the sample and hence the distance between the current electrodes is changed. The former method is equivalent to map the profile of the MI. Our results suggest that the local magnetization of grains and domain structure influences the MI, particularly in the first method. However, the length of the sample, which determines the demagnetizing effects, has a negligible influence on the magnitude and double peak behavior observed in the high frequency electrical transport.

We have measured the four probe electrical impedance ($Z(f,H)$ = $R(f,H)+jX(f,H)$, where $R$ is the ac resistance and $X$ is the inductive reactance, $X = \omega L$) of a bar shaped polycrystalline $La_{0.7}Sr_{0.3}MnO_3$ (LSMO) sample of cross sectional area 2.5x 2.5mm$^2$, using an Agilent 4285A LCR meter in the frequency range $f = \omega/2\pi$ = 0.5 MHz to 30 MHz, with a 50 mV ac excitation. An electromagnet was used to provide a variable dc magnetic field ($\mu_0H$) from -100 mT to +100 mT, where $H$ was applied parallel to the long axis of the sample, i.e., along the direction of the ac current. The electrical contacts were made with silver paste or silver-indium alloy. The electrodes for the current signal were made on the square faces at the two ends of the sample. The two linear voltage contact pads, each of about 100 µm wide, were placed symmetrically about the centre of the sample. The distance between the voltage probes ($l_v$) was changed after completing each set of measurements. Later, the same sample was cut equally at two ends after each set of measurements, while maintaining the distance between the symmetrically placed voltage probes constant, to reduce the sample length ($l_i$). The observed results were identical with same set of measurements performed on another bar shaped sample, cut from the parent polycrystalline disc.

Figure 1 shows the magnetic field dependence of the ac magnetoresistance, $\Delta R/R(0)$ (%) = [$R(H)$-$R(0)$]/$R(0)$ x 100, at 300 K for (a) $f <$ 15 MHz and (b) $f \geq$ 15 MHz respectively. The length between the voltage probes and current probes are $l_v$ = 9 mm and $l_i$ = 11 mm, respectively. The $\Delta R/R(0)$ at $f$ = 0.5 MHz decreases rapidly with increasing field and its field dependence closely matches with that of the dc MR, as can be seen in Fig. 1. However, its magnitude at $\mu_0 H$ = 100 mT is larger (≈26 %) than the dc value (< 1 %). As the frequency increases, the magnitude of $\Delta R/R(0)$ at $\mu_0 H$ = 100 mT markedly enhances, reaches a maximum value of ≈ 47 % at 5 MHz, and decreases to about 8 % at 30 MHz. The observed ac magnetoresistance of 47 % at 5 MHz in a small magnetic field of 100 mT is indeed remarkable compared to the smaller dc magnetoresistance (< 1 % at 100 mT).

Figures 1(c) and 1(d) show the magnetoreactance, $\Delta X/X(0)$(%) = [$X(H)$-$X(0)$]/$X(0)$ x 100 for $f <$ 15 MHz and $f \geq$ 15 MHz respectively for the same sample with $l_v$ = 9 mm and $l_i$ = 11 mm . In contrast to the $\Delta R/R(0)$, which is negative for all frequencies and shows a maximum around 5 MHz, $\Delta X/X(0)$ at 100 mT is maximum for $f$ = 0.5 MHz (≈ 25 %), and decreases at higher frequencies (< 1 % for $f$ = 20 MHz). More interestingly, the field dependence of $\Delta X/X(0)$ with increasing frequency is distinctively different from that of $\Delta R/R(0)$. The $\Delta X/X(0)$, which shows a single peak at the origin below 15 MHz, transforms into a valley with two peaks developing at $\pm H_p$ for $f > 15$ MHz. The amplitude of the double peak increases and $H_p$ shift upwards in the field with increasing frequency, and eventually changes sign from negative to positive. Only positive magnetoreactance in the entire field range is observed for $f = 30$ MHz.

In Fig. 2(a), we show the frequency dependence of $\Delta R/R(0)$ and $\Delta X/X(0)$ at 300 K for a sample with different distances between the voltage probes ($l_v$ (in mm) = 9, 6, 3, 2 and 1) where the distance between the current probes was kept constant ($l_i$ = 11 mm). When $l_v = 9$ mm, $\Delta R/R(0)$ initially increases, shows a maximum (= 47 %) around 5 MHz and then decreases at higher frequencies. The frequency dependence of $\Delta R/R(0)$ for other $l_v$s also shows a similar behavior. The $\Delta R/R(0)$ is nearly independent of frequency below 1 MHz, but shows a marked dependence on $l_v$ above 1 MHz. The magnitude of the maximum in $\Delta R/R(0)$ decreases with decreasing $l_v$ ($\approx$ 47 % for $l_v = 9$mm to $\approx 30$ % for $l_v = 1$ mm). Figure 2(b) shows the frequency dependence of $\Delta X/X(0)$ for various lengths between the voltage probes (the schematic is shown in the bottom inset of Fig. 2(b)). Here, the magnitudes of $\Delta X/X(0)$, particularly below 5 MHz, are dramatically affected by $l_v$. In the insets of Figs. 2(a) and 2(b), we show the frequency dependence of $\Delta R/R(0)$ and $\Delta X/X(0)$ respectively, for various length between the current probes ($l_i$ (in mm) = 11, 9, 6, and 3), for a fixed distance between the voltage probes ($l_v = 2.5$ mm). Interestingly, the frequency dependence does not show any remarkable variation for different lengths between the current probes.

Next we show how the distance between the voltage probes ($l_v$) influences the magnetic field dependence of impedance (discussed earlier in Fig. 1) at two selected frequencies, $f$ = 20 and 30 MHz, where the double peak structures are prominent.

Figures 3(a) and 3(b) show the field dependence of $\Delta R/R(0)$ at $f$ = 20 MHz, and 30 MHz respectively. We find that the magnitude of $\Delta R/R(0)$ systematically decreases with decreasing $l_v$. For example, the magnitudes of $\Delta R/R(0)$ at the maximum field for 20 (30) MHz is > 12 (7) % when $l_v$= 9mm, but it reduces to < 3 (1) % when $l_v$ = 1 mm. More interestingly, $\Delta X/X$ shows a striking difference between different $l_v$s. While $\Delta X/X(0)$ for $l_v$ = 9 mm at 20 MHz (Fig. 3(c)) shows a positive double peak at $H_p$ = ± 30 mT and negative values at higher fields, it shows an increasing tendency to be positive with decreasing $l_v$. It can be seen that $\Delta X/X(0)$ is positive over the entire field range for $l_v$ = 1 mm. However, the $\Delta X/X(0)$ at 30 MHz (Fig. 3(d)) for all $l_v$s remains positive and the magnitude of the peak decreases with decrease in $l_v$. We have also measured the dc magnetoresistance for various $l_v$, but the observed difference between the longest and the shortest $l_v$ was almost negligible within experimental errors.

Thus, our results clearly show that the observed MI effects are dependent not only on the frequency of the ac signal and field strength, but also on the length between the voltage probes of the sample. The ac impedance of a ferromagnetic sample is sensitive to the frequency and field dependence of the circular permeability, $\mu_\varphi(\omega, H)$. When the skin effect is weak ($\delta$ >> thickness of the sample), $Z = R_{dc} + j\omega L_i$, where $R_{dc}$ is the dc resistance and $L_i$ is the internal inductance of the sample. $L_i$ depends on the circular permeability through $L_i = G\mu_\varphi(\omega, H)$, where $G$ is the geometrical factor. Note that when the skin effect is strong, $Z \propto (1+j)\sqrt{(\omega\rho\mu_0\mu_\varphi/2)}$, the circular permeability affects both the ac resistance and the reactance.[2] Considering the complex nature of the permeability ($\mu_\phi = \mu_\phi' - j\mu_\phi''$), impedance in the weak skin effect regime under $H$ becomes $Z(\omega,H) = R_{dc}(H) + G\omega\mu_0\mu_\phi''(\omega,H) + jG\omega\mu_0\mu_\phi'(\omega,H)$. Thus, the reactance (*X*) depends on the real part of the circular permeability and the ac resistance (*R*) depends on the magnetic loss determined by the imaginary part of the permeability. Hence, the observed dependence on the distance between the voltage probes suggests that the high frequency electrical transport probes the local variations of circular permeability of ferromagnetic grains enclosed between the voltage probes. As $l_v$ increases, more number of ferromagnetic grains is sampled and as a result, average transverse permeability can increase in magnitude. Vasquez *et al.*[11] have found similar results in amorphous ferromagnetic FeCrSiBCuNb alloy wires, when one of the voltage probes was fixed at one end of the sample and another one was varied along the length of the sample. Local measurement of *B-H* loops indicated that the magnetization decreases and coercivity increases with decreasing distance between the voltage probes. It was suggested that the closure domain structure and domain wall pinning promotes magnetic hardness at the ends of the sample. The MI in our samples, surprisingly, does not depend on the length of the sample as long as the distance between the voltage probes is fixed and the magnetic field is applied along the length of the sample. This suggests that demagnetization effect plays less important role in the MI of bulk and thick manganites in axial dc magnetic field, which is opposite to the behavior found in other amorphous ferromagnetic alloys.[12] We found a coercive field of $H_C \approx 250$ mT at 10 K and about 40 mT at 300 K in our sample. The independence of MI on the length in our sample is possibly due to low value of magnetization (*M*) at small magnetic fields and hence much lower demagnetization field ($H_D = DM$ where D is the demagnetization factor) in manganites than in amorphous magnetic alloys at room temperature.

At low frequencies, both domain wall movement and domain rotation contributes to the permeability, but as the frequency increases domain wall motion is damped and hence the contribution of domain rotation to the permeability becomes more significant. The transition from a single to double peak in the field dependence of the magneto reactance signals an increasing role of the magnetization rotation within the domains.[5,6] The peak in $X$ occurs, when the dc bias field ($H$) balances the anisotropy field ($H_K$), and the amplitude of the peak increases with the magnitude of the permeability. The differences in the field depends of the magnetoreactance with changing distance between the voltage probes in LSMO sample at 30 MHz are possibly caused by the finer details of magnetic domain configurations in this sample and therefore needs further investigation.

In summary, our study indicates that while the ac magnetoresistance in radio frequency range in manganites is definitively higher than the dc magnetoresistance at the same field strength, the magnitude depends strongly on the measurement geometry, particularly the distance between voltage probes. Our study also indicates that the sign and the double peak features in the magnetoreactance are very sensitive to the frequency and the distance between the voltage probes. While detail understanding of the observed results requires knowledge of micromagnetism related to magnetization dynamics and domain structure, our results may set certain guidelines while considering the MI effect of these materials for practical applications. One of the possible practical implications of our investigation is to map the local electrical and magnetic response of grains in manganites in the radio frequency range using an X-Y scanning stage and understand correlations between local magnetization and magnetoresistance.[13]

R. M acknowledges National Research Foundation of Singapore for supporting this work through the Grant no : NRF-CRP-G-2007.

**Figure captions:**

**Fig. 1** (color online) Magnetic field dependence of the ac magnetoresistance ($\Delta R/R(0)$) for (a) $f <$ 15 MHz and (b) $f \geq$ 15 MHz. The ac magnetoreactance ($\Delta X/X(0)$) at various fixed frequencies for (c) $f <$ 15 MHz and (d) $f \geq$ 15 MHz. The numbers indicate the frequencies in MHz.

**Fig. 2** (color online) Frequency dependence of (a) $\Delta R/R(0)$ and (b) $\Delta X/X(0)$ for various distances between the voltage probes ($l_v$). The insets show the respective frequency dependence for various distances between the current probes ($l_i$).

**Fig. 3** (color online) Magnetic field dependence of the ac magnetoresistance $\Delta R/R(0)$) at (a) $f =$ 20 MHz and (b) 30 MHz for various $l_v$s. The magnetoreactance ($\Delta X/X(0)$) at (c) $f =$ 20 MHz and (d) 30 MHz are also shown. Note that the $\Delta X/X(0)$ for 30 MHz remains positive for all $l_v$s.

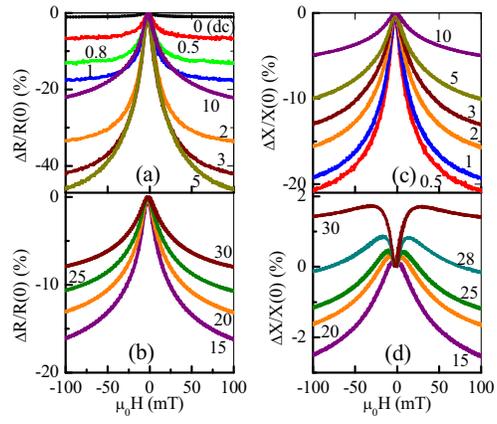

Fig. 1
Rebello *et al*.

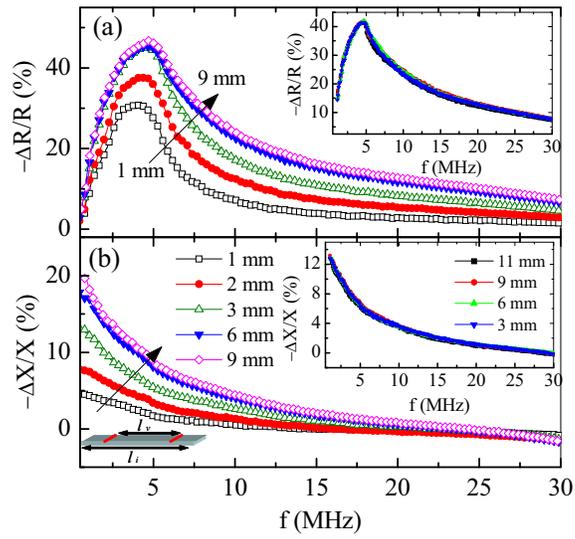

Fig. 2
Rebello *et al*.

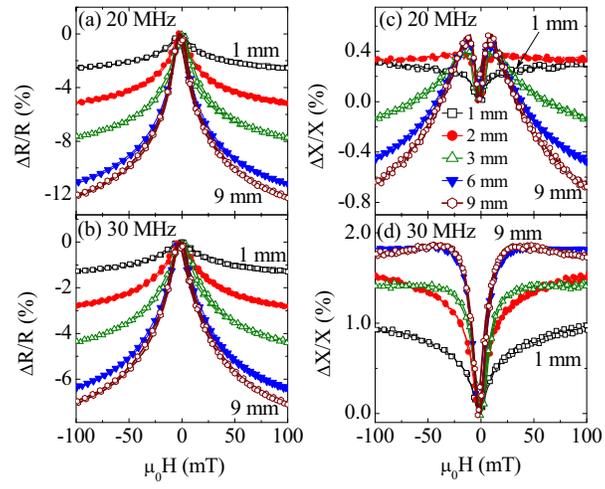

Fig. 3
Rebello *et al*.